\documentclass{emulateapj}

\usepackage{float}
\usepackage{amsmath}
\usepackage{epsfig,floatflt}
\usepackage{subfigure}

\begin{document}

\title{Astro-ph communication: \\ Simulations of the \emph{WMAP} Internal
  Linear Combination sky map}

\author{H.\ K.\ Eriksen\altaffilmark{1}, A.\ J.\
  Banday\altaffilmark{2}, K.\ M.\ G\'orski\altaffilmark{3} and P.\ B.\
  Lilje\altaffilmark{4}}

\altaffiltext{1}{Institute of Theoretical Astrophysics, University of
Oslo, P.O.\ Box 1029 Blindern, N-0315 Oslo, Norway; Centre of
Mathematics for Applications, University of Oslo, P.O.\ Box 1053
Blindern, N-0316 Oslo; JPL, M/S 169/327, 4800 Oak Grove Drive,
Pasadena CA 91109; California Institute of Technology, Pasadena, CA
91125; email: h.k.k.eriksen@astro.uio.no}

\altaffiltext{2}{Max-Planck-Institut f\"ur Astrophysik, Karl-Schwarzschild-Str.\
1, Postfach 1317,D-85741 Garching bei M\"unchen, Germany; email:
banday@MPA-Garching.MPG.DE} 

\altaffiltext{3}{JPL, M/S 169/327, 4800 Oak Grove Drive, Pasadena CA
  91109; Warsaw University Observatory, Aleje Ujazdowskie 4, 00-478
  Warszawa, Poland; California Institute of Technology, Pasadena, CA
  91125; email: Krzysztof.M.Gorski@jpl.nasa.gov}

\altaffiltext{4}{Institute of Theoretical Astrophysics, University of
Oslo, P.O.\ Box 1029 Blindern, N-0315 Oslo, Norway; Centre of
Mathematics for Applications, University of Oslo, P.O.\ Box 1053
Blindern, N-0316 Oslo; email: per.lilje@astro.uio.no}

\begin{abstract}
Following this astro-ph communication, we release a set of 10\,000
\emph{WMAP} ILC simulations, produced as described by
\citet{eriksen:2004}. We strongly encourage that an analysis of these
simulations accompanies any scientific analysis of the observed ILC
map, in order to assess the stability of the results. For examples of
such analyses, we refer the reader to our original paper. In
particular, Figure 1c) and d) show the bias and uncertainty induced by
the ILC method, and Figure 7 shows a sobering examples of a low-$\ell$
analysis; the reconstructed quadrupole-octopole alignment is plotted
against the true input alignment, and the correlation is
underwhelming. It can hardly be emphasized too strongly -- the heavily
processed full-sky \emph{WMAP} maps are not reliable for quantitative
analysis, neither at large nor small scales, and they should only, at
best, be used as supportive evidence. We hope that these simulations
will reduce some of the confusion regarding these maps that currently
exists in the community.
\end{abstract}


\maketitle

The main data product of the first-year \emph{WMAP} data release
\citep{bennett:2003a} was a set of ten sky maps covering frequencies
between 23 and 94 GHz. These sky maps formed the basis of both
cosmological and astrophysical analyses, and unprecedented constraints
on both the CMB power spectrum and cosmological parameters were
obtained.

For public convenience, a set of template-corrected sky maps were also
provided by the \emph{WMAP} team on the
LAMBDA\footnote{http://lambda.gsfc.nasa.gov} web-site. For most
cosmological applications, these constitute the data set of choice
(coupled with a proper sky cut), since they have both well-understood
noise properties and fairly low levels of residual foreground
contamination. 

The \emph{WMAP} team also released a full-sky Internal Linear Combination map
\citep{bennett:2003b} intended for visualization purposes only. Its
complicated noise and foreground properties (and mixing of the two)
make it most unreliable for quantitative analysis. However, since the
Galactic plane is not readily visible, it is useful for public
display. The \emph{WMAP} team explicitly warns against using it for CMB
analysis. 

For some applications, full-sky maps are highly desirable due to
algorithmic limitations. In particular, methods that operate in
spherical harmonics space are greatly simplified if no sky cut is
imposed in the analysis. For this reason, the ILC map, which by eye
looks almost free of foreground residuals, has been extensively used
for scientific purposes -- despite the fact that there are strong (and
difficult to quantify) residual foregrounds present in the map. 

This provided the motivation for our first analysis of the ILC map,
which is detailed by \citet{eriksen:2004}. In this study we found that
the ILC map is indeed highly contaminated by residual foregrounds, and
in particular, that the low-$\ell$ components, which have received the
most attention so far, are highly unstable under the ILC cleaning
operation. A sobering example is the analysis of reconstructed versus
input quadrupole-octopole alignment, which is summarized in Figure 7
of the above paper: Clearly, the preferred directions of the
low-$\ell$ components are more or less randomized by the residual
foregrounds present in the ILC map. 

During that analysis, we also found that the weights used to form the
official \emph{WMAP} ILC sky map were sub-optimal: Lower variances could be
found by weighing the frequency bands slightly differently. This was
due to poor convergence in the non-linear search algorithm used by the
\emph{WMAP} team, and was thus purely an algorithmic effect. Rather than
relying on non-linear searches, we therefore solved the problem using
Lagrange multipliers. However, since neither our map nor the one
produced by the \emph{WMAP} team is intended for quantitative analysis, the
difference did not matter. But the two maps should not be considered
independent, and the fact that they agree to some extent should not be
taken as evidence of low foreground contamination.

Given all the attention these maps have received, we believe it is
useful to make a set of ILC simulations publicly available. These
simulations are produced by adding foreground templates and
channel-specific noise to random CMB realizations, and running these
``semi-realistic'' simulations through the ILC algorithm. By comparing
the input CMB realization to the reconstructed realization with some
statistic of choice, one gains some intuition of the errors introduced
by the ILC method. However, we point out that the errors are likely to
be \emph{underestimated}, because of the constant spectral indices
implied by the foreground templates, and also because the template
coefficients are estimated from the high-latitude sky only, and this
leads to underestimation of the low-latitude foreground levels. These
simulations should therefore be interpreted as setting a lower limit
on the residuals present in the observed ILC map.

It is also worth noticing that in order to estimate the errors in the
ILC map, it is not sufficient to simply add scaled foreground
templates to CMB realizations. For the WMAP ILC map, individual
weights are computed for 12 disjoint regions, and the interplay
between noise and foregrounds within each of these regions makes it
very hard to predict the full-sky results. End-to-end simulations such
as the ones provided here is should be studied to understand the
effects of the ILC method on the statistic of interest.

A set of 10\,000 ILC simulations may be freely downloaded here:
\begin{itemize}
\item http://www.astro.uio.no/$\sim$hke/\\
\hspace*{1cm}  /cmbdata/LILC\_sim\_input.tar.gz
\item http://www.astro.uio.no/$\sim$hke/\\
\hspace*{1cm}  /cmbdata/LILC\_rec\_input.tar.gz 
\end{itemize}
The first file contains the input CMB realizations, and the second
file contains the reconstructed ILC maps. 

For storage reasons, these two files contain the spherical harmonics
coefficients $a_{\ell m}$ (in standard HEALPix format) up to $\ell=50$
for each map, which hopefully should be sufficient for more low-$\ell$
analyses. The 10\,001.\ realization in the reconstruction set contains
the actual observed LILC map. Other formats or $\ell$-ranges may be
obtained by sending an email to HKE.

The spherical harmonics coefficients have been multiplied with the
response function of a Gaussian beam of $1^{\circ}$ FWHM and a pixel
window function of corresponding to a HEALPix resolution parameter
$N_{\textrm{side}} = 512$. Thus, if a sky map is desired, as opposed
to spherical harmonics coefficients, the data as provided here must be
convolved with a beam of at least $8^{\circ}$ FWHM to have a properly
bandwidth limited function at $\ell_{\textrm{max}}=50$. If the
functions are projected on grids with lower resolution than
$N_{\textrm{side}}=512$, one also must convolve with the appropriate
pixel window. 

We also point out that if one wants to study the low-$\ell$ components
of the \emph{WMAP} data, the reconstruction method described by
\citet{bielewicz:2004} applied to the template-corrected sky maps is
significantly more accurate than the ILC method. The uncertainties
induced by the cut sky are considerably smaller than the
foreground-induced uncertainties in the ILC maps (see, e.g., Figure 1
of Bielewicz et al.\ 2005).

We hope that this set of simulations will prove useful for future
analyses, and that they may help reducing the confusion caused by
these ``foreground-cleaned'' full-sky maps.

\end{document}